\documentclass[8pt]{iopart}
\usepackage{iopams} 
\usepackage[latin1]{inputenc}
\usepackage{graphicx}
\usepackage{url}

\begin{document} 
 
\title[Spontaneous emission -- graphene waveguides -- surface plasmon polaritons]{Spontaneous emission in plasmonic graphene subwavelength wires of arbitrary sections}
\author{Mauro Cuevas} 
\address{Consejo Nacional de
Investigaciones Cient\'ificas y T\'ecnicas (CONICET) and Facultad de Ingenier\'ia y Tecnolog\'ia Inform\'atica, Universidad de Belgrano,
Villanueva 1324, C1426BMJ, Buenos Aires, Argentina}
\address{Grupo de Electromagnetismo Aplicado, Departamento de F\'isica, FCEN, Universidad de Buenos Aires,  Ciudad Universitaria,
Pabell\'on I, C1428EHA, Buenos Aires, Argentina}
\ead{cuevas@df.uba.ar}

\begin{abstract}
We present a theoretical study of the spontaneous emission of a line dipole source embedded  in a graphene--coated subwavelength wire of arbitrary shape. The modification of the emission and the radiation efficiencies are calculated by means of a rigorous electromagnetic method based on Green's second identity. Enhancement of these efficiencies is observed when the emission frequency coincides with one of the plasmonic resonance frequencies of the wire. The relevance of the dipole emitter position and the dipole moment orientation are evaluated. We present calculations of the near--field distribution for different frequencies which reveal the multipolar order of the plasmonic resonances. 
%
%
\end{abstract}



\maketitle
\ioptwocol

\section{Introduction} 

It is known that the  spontaneous emission rate of an excited atom or molecule is not an intrinsic attribute,  but it depends on the surrounding environment, being a property that has been exploited to improve the efficiency of current light control devices, such as photonic band gaps \cite{yablonovitch} and highly efficient single photon sources  \cite{SPS}.  
As pointed out by Purcell \cite{purcell}, the spontaneous emission rate is proportional to the local density of electromagnetic states, particularly, to the local electromagnetic field confinement  %
 in the vicinity  of the molecule. 
In addition, surface plasmons (SPs) have attracted interest due to their unique property to confine  a great amount of electromagnetic energy  at subwavelength scales and 
for providing strong coupling of light to emitter systems. 
Enhanced emission rate due to SP excitations in a variety of metal nanostructures  such as uniform planar microcavities, metal nanoparticles and metal nanowires, have been the subject  of many theoretical and experimental investigations \cite{vukovic,nanotriangulos,rogobete,wire1}. 

The recent advent of graphene has  met the need of SPs at terahertz frequencies, since it offers relatively low loss, high confinement and good tunability \cite{jablan,Xia}. Significant progress made in nanoscale fabrication and an extensive wealth of theoretical analysis have allowed possible a wide range  of  applications based on the interaction between graphene and electromagnetic radiation via SP mechanisms, including plasmonic signal
processing \cite{procesing}, sensing \cite{velichko,s2}, quantum optics and nonlinear photonics \cite{maier_nature,garciadeabajo}. 

SP--induced modifications of the spontaneous emission have become the focus of particular attention in different graphene based structures, such as infinite graphene monolayers \cite{grafeno4}, ribbons or nanometer sized disks \cite{ribons,disk},  double--layer graphene waveguides and  single--walled carbon nanotubes \cite{cuevas1,martin_moreno}.  
Controllable assembly of single atom, molecules, quantum dots and nanoparticles on  graphene platforms \cite{encapsulado1,encapsulado2,francisco2} opens up possibilities for practical optoelectronic applications involving graphene hybrid structures. 

In a recent work \cite{cuevas2}, we have revealed that
the interplay between the optical emitter and SP excitations on the graphene layer 
 strongly influences both the spontaneous emission as well as the radiation characteristics of an optical emitter into a graphene coated wire of circular cross section. Since the electric field of the excited SPs  reaches similar values inside and outside the graphene wire \cite{CRD}, and taking into account new possibilities of encapsulating single atoms, molecules and compounds into graphene wires \cite{encapsulado1,encapsulado2}, and the fact that, as a result of the van der Waals force, a graphene sheet can be tightly coated on a fiber surface \cite{coated}, in that work we deemed appropriate to consider the case in which the optical emitter is localized inside the wire.   

Our primary motivation for this work is to extend 
the study realized in \cite{cuevas2} to the case of wires with an arbitrary cross section.  
To do this, the calculations have been made using the Green function surface integral method (GSIM) \cite{sanchezgil,GSIM,josab} which enables to solve the scattering problem for structures with a complex shape.  The GSIM  is well established for electromagnetic scattering problems and it has been applied to deal with a wide variety of problems such as  the scattered field produced by an incident wave onto a random grating  \cite{m}, the interaction of light with particles of arbitrary shapes \cite{valencia}, the scattered field by metal nanostrip resonators placed close to a surface \cite{soresen2} and the electromagnetic response of a metamaterial surface with a localized defect \cite{PLA}.

We have represented the graphene by an  infinitesimally thin, two--sided layer with a frequency--dependent surface conductivity as used in Ref. \cite{hanson} where the electromagnetic field due to an electric current  is obtained in terms of dyadic Green$'$s functions represented as Sommerfeld integrals. We believe that this approach is particularly appropriate and, in our opinion, even superior to other alternatives where the graphene is modeled as a layer of certain thickness and with an effective refractive index since the latter case  fails to explain some of the results in  remarkable optical experiments \cite{merano}.

Using the GSIM in this paper we explore 
 the effects that the departure from a circular cross section 
has on the emission and the radiation efficiencies of a single emitter placed inside the wire. In particular,  we focus our study on wires whose  cross section has  a  super--elliptical form, a shape intermediate between ellipse and rectangle \cite{gelis}.

This paper is organized as follows. First, in Section \ref{teoria} we provide the expressions for the electromagnetic field  generated   by a line dipole source coupled to a  graphene--coated wire. The source  is located at an arbitrary position and with an arbitrary orientation inside the wire. 
 The scattered field is expressed in terms of two unknown source functions evaluated on the graphene layer, one related to the field interior to the wire and the other related to its normal derivative. In Section \ref{resultados} we validate the numerical results by comparing them with analytical calculations in the  case of a circular wire \cite{cuevas2}. To explore the effect that the deviation from the circular geometry has on the emission and the radiation spectra of the single emitter, we consider graphene coated wires of quasi--rectangular shapes. Finally,  concluding remarks are provided in Section \ref{conclusiones}. 
The Gaussian system of units is used
 and an $\mbox{exp}(-i\, \omega\, t)$ time--dependence is implicit throughout the paper, with $\omega$ as the angular frequency, $t$ as the time, and $i=\sqrt{-1}$. The symbols Re and Im are respectively used for denoting the real and imaginary parts of a complex quantity.

\section{Theory} \label{teoria} 

\begin{figure}
\centering
\resizebox{0.40\textwidth}{!}
{\includegraphics{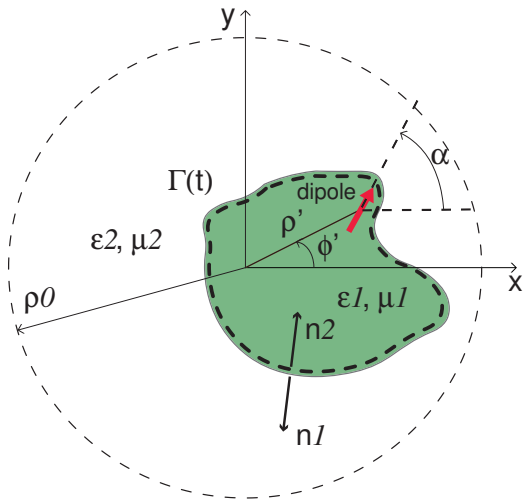}}
\caption{\label{fig:epsart} Schematic illustration of the system. An optical dipole emitter is inside a graphene--coated dielectric cylinder. The wire ($\varepsilon_1$, $\mu_1$ and surface conductivity $\sigma$) is embedded in a transparent medium with constitutive parameters $\varepsilon_2$, $\mu_2$.}\label{sistema}
\end{figure}

We consider the scattering problem of a line dipole source radiating inside a graphene coated wire cylinder (figure 1). We assume that both the cylinder and the dipole line axis lie along the $\hat{z}$ axis. The cross section of the cavity wire is defined by a planar curve described by the vector valued function $\Gamma(t)=x(t)\hat{x}+y(t)\hat{y}$ and the wire substrate is characterized by the electric permittivity $\varepsilon_1$ and the magnetic  permeability $\mu_1$. The wire is embedded in a transparent medium with electric permittivity $\varepsilon_2$ and magnetic permeability $\mu_2$.    
The graphene layer is considered here as an infinitesimally thin, local and isotropic two--sided layer with frequency--dependent surface conductivity $\sigma(\omega)$ given by the Kubo formula \cite{falko}, which can be read as  $\sigma= \sigma^{intra}+\sigma^{inter}$, with the intraband and interband contributions being
\begin{equation} \label{intra}
\sigma^{intra}(\omega)= \frac{2i e^2 k_B T}{\pi \hbar (\omega+i\gamma_c)} \mbox{ln}\left[2 \mbox{cosh}(\mu_c/2 k_B T)\right],
\end{equation}  
\begin{eqnarray} \label{inter}
\sigma^{inter}(\omega)= \frac{e^2}{\hbar} \Bigg\{   \frac{1}{2}+\frac{1}{\pi}\mbox{arctan}\left[(\omega-2\mu_c)/2k_BT\right]-\nonumber \\
   \frac{i}{2\pi}\mbox{ln}\left[\frac{(\omega+2\mu_c)^2}{(\omega-2\mu_c)^2+(2k_BT)^2}\right] \Bigg\},
\end{eqnarray}  
where $\mu_c$ is the chemical potential (controlled with the help of a gate voltage), $\gamma_c$ the carriers scattering rate, $e$ the electron charge, $k_B$ the Boltzmann constant and $\hbar$ the reduced Planck constant.
The intraband contribution dominates for large doping $\mu_c<<k_BT$ and it is a generalization of the Drude model for the case of arbitrary band structure, whereas the interband contribution dominates for large frequencies $\hbar \omega \geq \mu_c$. When the line source with a dipole moment $\vec{p}=p [\cos \alpha \, \hat{x}+ \sin \alpha\,\hat{y}]$ is placed inside the plasmonic cavity wire ($\alpha$ is the angle between the dipole moment and the $\hat{x}$ axis), the magnetic field is along the $\hat{z}$ axis ($\vec{H}(\vec{r})=\varphi(\vec{r}) \, \hat{z}$). The wave equation for $\varphi(\vec{r})$ reads 
%
%
%
\begin{eqnarray}\label{A}
\nabla^2 \varphi_j(\vec{r})+k_j^2 \varphi_j(\vec{r}) = g_j(\vec{r},\vec{r}_s),
\end{eqnarray} 
where subscripts $j=1,\,2$ is used to denote the internal region (medium 1) and the exterior region (medium 2)  to the boundary wire, respectively, $k_j=k_0 \sqrt{\varepsilon_j\,\mu_j}$, 
$k_0=\omega/c$ is the modulus of the photon wave vector in vacuum, $\omega$ is the angular frequency, $c$ is the vacuum speed of light, $g_1(\vec{r},\vec{r}_s)=-4\,\pi\, i k_0 \,\vec{p} \times\,\nabla \delta(\vec{r}-\vec{r}_s)$, $g_2(\vec{r},\vec{r}_s)=0$ and $\vec{r}_s$ denotes the position of the line source.  
To solve Eq. (\ref{A}), we transform it into a boundary integral equation using the GSIM as explained in \cite{GSIM,josab}. Using Eq. (\ref{A}) in the interior region and separating the total field in this region into contributions from the primary field emitted by the dipole and scattered field $\varphi_1(\vec{r})=\varphi_{inc}(\vec{r})+\varphi_s(\vec{r})$ we obtain 
%
\begin{eqnarray}\label{phi1}
\varphi_1(\vec{r})=\varphi_{inc}(\vec{r})+\nonumber\\
\frac{1}{4\pi}\int_{\Gamma}\,\left[\frac{\partial\,G_1(\vec{r},\vec{r}')}{\partial n'}\varphi_1(\vec{r}')-G_1(\vec{r},\vec{r}') \frac{\partial\,\varphi_1(\vec{r}')}{\partial n'}\right]\,ds' ,
\end{eqnarray} 
where $\vec{r}'$ is a point on the boundary $\Gamma(t)$ with arc element $ds'$, the derivative  $\frac{\partial}{\partial n'}$  along the normal to the interface at  $\vec{r}'$ is directed from the medium 2 to the medium 1 ($n'=n_2=-n_1$), and $G_1(\vec{r},\vec{r}')$ is the Green function of Eq. (\ref{A}) in the interior region 
\begin{eqnarray}\label{funcion de green}
G_1(\vec{r},\vec{r}')=i \pi H_0^{(1)}(|\vec{r}-\vec{r}'|), 
\end{eqnarray}
where $H_0^{(1)}$ is the 0th Hankel functions of the first kind, and 
\begin{eqnarray}\label{inc}
\varphi_{inc}(\vec{r})=i k_0 \hat{z}\, (\vec{p} \times \nabla)  G_1(\vec{r},\vec{r}_s).
\end{eqnarray}
Similarly, outside the wire region the field  is 
\begin{eqnarray}\label{phi2}
\varphi_2(\vec{r})=\\   \nonumber
-\frac{1}{4\pi}\int_{\Gamma}\,\left[\frac{\partial\,G_2(\vec{r},\vec{r}')}{\partial n'}\varphi_2(\vec{r}')-G_2(\vec{r},\vec{r}') \frac{\partial\,\varphi_2(\vec{r}')}{\partial n'}\right]\,ds' ,
\end{eqnarray} 
where $G_2(\vec{r},\vec{r}')$ is the Green function in the exterior region to the wire. From Eqs. (\ref{phi1}) and (\ref{phi2}), the total field in regions 1 and 2 are completely determined by the boundary values of the field and its normal derivative. By allowing the point of observation $\vec{r}$ to approach the surface in Eqs. (\ref{phi1}) and (\ref{phi2}), we obtain a pair of coupled integral equations with four unknown functions: the values of $\varphi_j$ and $\partial \varphi_j / \partial n$, $j=1,\,2$, at the $\Gamma$ boundary.      
The electromagnetic boundary conditions at $\Gamma$,
\begin{eqnarray}\label{cc1}
\frac{1}{\varepsilon_1}\frac{\partial \varphi_1}{\partial n_1}=\frac{1}{\varepsilon_2}\frac{\partial \varphi_2}{\partial n_1},
\end{eqnarray} 
and 
\begin{eqnarray}\label{cc2}
\varphi_2-\varphi_1= 
\frac{4 \pi \sigma}{c k_0 \varepsilon_1}i \frac{\partial \varphi_1 }{\partial   n_1}, 
\end{eqnarray} 
%
provides two additional relationships between the fields and their normal derivatives at the boundary of the wire,  allowing us to express $\varphi_2$ and $\partial \varphi_2 / \partial n$ in terms of  $\varphi_1$ and $\partial \varphi_1 / \partial n$. To find these functions we convert  the system of integral equations into matrix equations which are solved numerically  (see \cite{josab} and references therein). 
Once the functions  $\varphi_1$ and $\partial \varphi_1 / \partial n$ are determined, the scattered field, given by Eqs. (\ref{phi1}) and (\ref{phi2}) can be calculated at every point in the interior  and exterior  regions.  The time--averaged  power emitted can be calculated from the integral of the normal component of the complex Poynting vector flux through the inner side of the boundary wire (see Figure \ref{sistema})
\begin{eqnarray}\label{P}
P= L\oint_{\Gamma} \mbox{Re} \left\{\vec{S_1}(\vec{r}')\cdot\hat{\mbox{n}}_1 \right\} ds',
\end{eqnarray}
where 
\begin{eqnarray}\label{S}
\vec{S_1}(\vec{r}')= \frac{c}{8\pi} \left(-\frac{i}{k_0 \varepsilon_1}\right) \varphi_1^\ast(\vec{r}') 
\left[\frac{\partial \varphi_1(\vec{r}')}{\partial n_1}\right].
\end{eqnarray}
Similarly, the time--averaged radiative power  can be  evaluated by calculating the complex Poynting vector flux through an imaginary cylinder of length $L$ and radius $\rho_0$ that encloses the cavity wire  (see Figure \ref{sistema})
\begin{eqnarray}\label{Pr}
P_{s}=\\ \nonumber
\frac{ \rho_0\,L\,c^2}{8 \pi \omega \varepsilon_2} \int_0^{2 \pi} \mbox{Re} \left\{ -i\, \left[\varphi_2(\rho_0,\phi)\right]^\ast \frac{\partial \varphi_2(\rho_0,\phi)}{\partial \rho} \right\} d\phi.
\end{eqnarray}
In the far--field region the calculation of the scattered fields given by  Eq. (\ref{phi2}) can be greatly simplified using the asymptotic expansion of the Hankel function for large argument \cite{abramowitz}. After some algebraic manipulation, we obtain \cite{josab}
\begin{eqnarray}\label{Pr2}
P_{s}=\frac{ L\,c^2}{64 \pi^2 \omega \varepsilon_2} \int_0^{2 \pi} |F(\phi)|^2 d\phi,
\end{eqnarray}
where 
\begin{equation}\label{Pr3}
\begin{array}{ll}
F(\phi)=\\
\oint_{\Gamma(t)} i k_2 \left([-y'(t) \cos(\phi)+x'(t) \sin(\phi)] \varphi_2(t)+\frac{\partial \varphi_2(t)}{\partial n_1}\right) \\
& \\
\times e^{-ik_1(x(t) \cos(\phi)+y(t)\sin(\phi)} dt.
\end{array}
\end{equation}
We define the normalized spontaneous  emission rate  $F$ as the ratio between the power emitted by the dipole, given by Eq. (\ref{P}), and the power emitted by the same dipole  embedded in an unbounded medium 1. 
In a similar way,
 the radiative efficiency $F_{s}$ is defined as the ratio between the power radiated by the dipole, given by Eq. (\ref{Pr2}), and the power  emitted by the dipole in the unbounded medium 1. 
\section{Results}\label{resultados}

\begin{figure}
\centering
\resizebox{0.5\textwidth}{!}
{\includegraphics{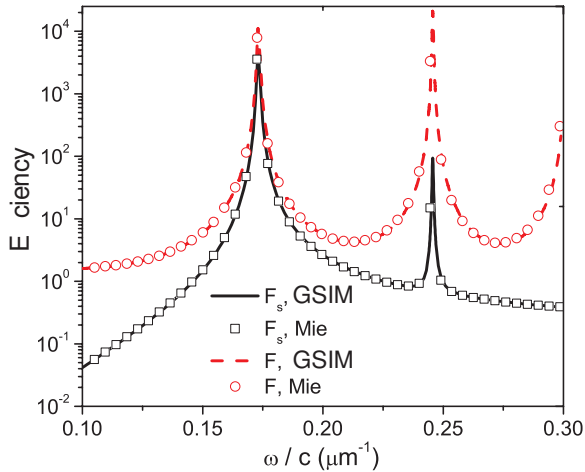}}
\caption{\label{fig:epsart} Comparison between the integral and the multipolar Mie formalisms results for the emission and the radiation efficiency, $F$ and $F_{s}$, of a graphene coated dielectric cylinder with a radius  $a=0.5\mu$m, constitutive parameters  $\varepsilon_1=3.9$ and $\mu_1=1$ in a vacuum. The graphene parameters are $\mu_c=0.5$eV, $T = 300$ K and  $\gamma_c=0.1$meV. The emitter is localized at $\rho'=0.4\mu$m with its dipole moment oriented in the $+\hat{x}$ direction ($\alpha=0$). 
}\label{comparacion}
\end{figure}

We have firstly  validated the numerical results by comparing them to analytical calculations in the case of a circular dielectric cylinder tightly coated with a graphene layer. 
The core (radius $a=0.5\mu$m) is made of a transparent material ($\varepsilon_1 = 3.9$, $\mu_1=1$)  and is embedded in vacuum ($\mu_2 = \varepsilon_2 = 1$). We used Kubo parameters $\mu_c=0.5$eV, $\gamma_c=0.1$meV, $T=300$K, emission frequencies in the range between $0.1\mu$m$^{-1}$ ($5$ THz or wavelength $\lambda=60\mu$m) and $0.3\mu$m$^{-1}$ ($15$ THz or wavelength $\lambda=20\mu$m),  and the emitter is localized at $\rho'=0.4\mu$m.  
Figure \ref{comparacion} shows both $F$ and $F_{s}$ efficiencies obtained with the integral formalism described in this paper (solid and dashed curves) and with the analytical results obtained using solution for the scattered fields in the form of infinite series of cylindrical harmonics (squares and circles) sketched in Ref. \cite{cuevas2}.  
A good agreement between both formalisms is observed in this Figure.  
The spectral position of the multipolar plasmon resonances, at a frequency near $0.17\mu$m$^{-1}$ for the dipolar resonance and near $0.24\mu$m$^{-1}$ for the quadrupolar resonance, also agree well with those obtained 
 from the quasistatic  approximation for which the stationary plasmonic mode condition is fulfilled \cite{CRD}. This condition asserts that 
a plasmonic mode of a graphene--coated circular cylinder accommodates, along the cylinder circumference, an integer number of oscillation periods of the propagating surface plasmon corresponding to the flat graphene sheet \cite{CRD}. This surface plasmon has been obtained in \cite{hanson}, and experimentally confirmed based on analysis of existing data by Merano \cite{merano2}, as a proper mode propagating with its electric and magnetic fields  decaying exponentially away from the plane graphene sheet in two adjacent regions.

\begin{figure}
\centering
\resizebox{0.40\textwidth}{!}
{\includegraphics{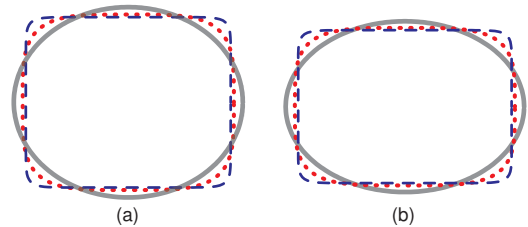}}
\caption{\label{fig:epsart} Quasi--square ($a = b = 1$) (a) and quasi--rectangular ($a=1.25, \, b=1$) (b) wire shapes delimited by equal--perimeter planar curves described by Eq. (\ref{superelipse}). $n=4$ (doted line), $n=12$ (dashed line). As a reference, a circumference (a) and an ellipse (b) whose lengths are equal to the polygon perimeter are also shown (continuous line).} \label{perfiles}
\end{figure}

In order to explore the effects that the departure from the circular geometry has on the spectrum of a dipole emitter localized inside the wire, we now use the GSIM to calculate the emission and the radiation decay rates in graphene 
coated wires of super--elliptical shapes delimited by planar curves which, in polar parametrization, are described by  \cite{gelis}
\begin{eqnarray}\label{superelipse}
\rho(\phi)= \frac{h}{[|\cos(\phi) / a|^n+|\sin(\phi) / b|^n]^{1/n} }.
\end{eqnarray}
%
By varying the values of the parameters $a,\, b,\,\mbox{and}\, n$, a wide range of natural and engineered shapes can be modeled by this parametrization. In fact, the curve described by Eq. (\ref{superelipse}) approaches a rectangle when $n$ increases and it  degenerates into an ellipse  when $n = 2$. 
In our simulations we used quasi--square and quasi--rectangular wire sections such as those shown in Fig. \ref{perfiles}. As a reference, a circumference (ellipse) whose length is equal to the square (rectangle) perimeters is also shown. To  compare the results with those obtained in the circular case, 
all delimiting curves in this figure have the same perimeter as that of the circular wire in Figure \ref{comparacion}, a condition obtained by properly selecting the scaling factor $h$ in Eq. (\ref{superelipse}). 

\begin{figure}
\centering
\resizebox{0.48\textwidth}{!}
{\includegraphics{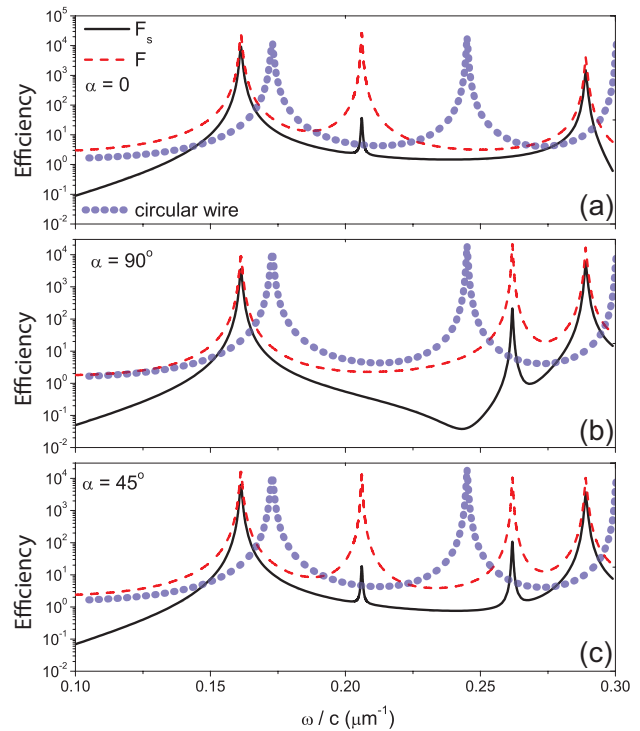}}
\caption{\label{fig:epsart} Efficiency (per unit length) curves,  calculated for $\mu_c=0.5$eV, $T = 300$ K, $\gamma_c=0.1$meV, $\varepsilon_1=3.9$, $\mu_1=1$ and $\varepsilon_2=1$, $\mu_2=1$. The wire shape is delimited by Eq. \ref{superelipse} with $a=b=0.447$ and $n=12$. The scaling factor $h$ is chosen so that the wire has the same perimeter as that of the circle in Figure \ref{comparacion}.  
The emitter is  localized at $\rho'=0.347\mu$m, $\phi'=180^\circ$ with its dipole moment oriented in the $+\hat{x}$ direction ($\alpha=0$) (a), in the $+\hat{y}$ direction ($\alpha=90^\circ$) (b), and $\alpha=45^\circ$ (c). The emission curve corresponding to the circular wire  is given as a reference.} \label{potencia_cuadrado_12}
\end{figure}

In  Figure \ref{potencia_cuadrado_12} we plot the normalized spontaneous emission and the scattering decay rates $F$ and $F_s$ for an emitter positioned on the $\hat{x}$ axis inside a graphene quasi--square wire with $a=b$ and $n=12$. To illustrate the effects of varying the $\alpha$ orientation angle, in Fig. \ref{potencia_cuadrado_12} we show these curves for different $\alpha$ values,  $\alpha= 0,\, 90^\circ,\,\mbox{and}\,45^\circ$, when the source is placed at $\rho'=0.347\mu$m and $\phi'=180^\circ$ (at $0.1\mu$m from the left of the  wire).  The emission spectrum corresponding to the circular case ($n=2$) is also given as a
reference. 
We observe that both $F$ and $F_s$ are enhanced at a frequency near $0.16\mu$m$^{-1}$ corresponding to the dipolar plasmon resonance of the graphene quasi--square wire.  The appearance of this peak does not depend on the orientation angle $\alpha$ 
and it is shifted to lower frequencies with respect to the peak corresponding to the excitation of the dipolar resonance in a circular wire (plotted with a dotted curve in Figure \ref{potencia_cuadrado_12}) for which the stationary mode condition is fulfilled.
The break of the rotational symmetry  of the wire section introduces a two--dimensional anisotropy in the emission and radiation spectrum, particularly evident in a frequency splitting of the quadrupolar resonance.
\begin{figure}
\centering
\includegraphics[width=\linewidth]{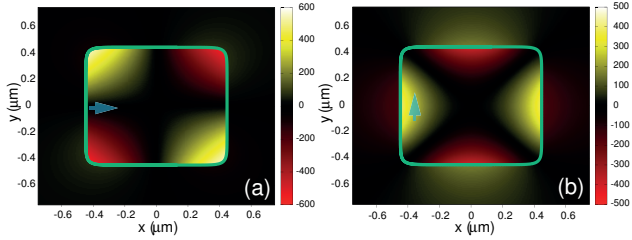}
\caption{\label{fig:epsart} Map of the scattered magnetic field $\varphi_s(\rho,\phi)$ at a fixed time for plasmon resonances of the wire considered in Figure \ref{potencia_cuadrado_12}. Red, negative values, yellow, positive  values. The frequency emission is $\omega/c = 0.206 \mu$m$^{-1}$ (a), $\omega/c = 0.262 \mu$m$^{-1}$ (b). The dipole emitter is indicated with an arrow.   
All parameters are the same as in Figure \ref{potencia_cuadrado_12}.} \label{cuadrupolo_a1b1}
\end{figure}
In Figure \ref{potencia_cuadrado_12}c we observe this splitting    
when the dipole axis forms an angle $\alpha=45^\circ$ with the $+\hat{x}$ axis, which in the circular case occurs at a frequency near $0.24 \mu$m$^{-1}$  and that in the quasi--square case is split
into a peak near $0.20\mu$m$^{-1}$ 
and another peak near $0.26\mu$m$^{-1}$. 
The first peak appears when  the dipole orientation $\alpha=0$ and the second peak appears when the dipole orientation $\alpha=90^\circ$, as is indicated in Figure \ref{potencia_cuadrado_12}a and \ref{potencia_cuadrado_12}b due to the fact that both resonances are decoupled for dipole orientations parallel to either $\hat{x}$ or $\hat{y}$ axis and that the first (respectively second) peak is absent when the dipole moment is oriented along the $\hat{y}$ (respectively the $\hat{x}$) axis. 
These resonances  are  appreciated in the near field, as shown in Fig. \ref{cuadrupolo_a1b1} where we plot the spatial distribution of the near scattered magnetic field $\varphi_s(\rho,\phi)=\varphi_1(\rho,\phi)-\varphi_{inc}(\rho,\phi)$ for $\omega/c=0.206\mu$m$^{-1}$   with the dipole orientation along the $\hat{x}$ axis ($\alpha=0$) and $\omega/c=0.262\mu$m$^{-1}$ with the dipole orientation along the $\hat{y}$ axis ($\alpha=90^\circ$). In the first case,  the field appears enhanced at the corners of the square wire, while in the second case the field is enhanced at the adjacent sides. We have verified (not shown in Fig. \ref{cuadrupolo_a1b1}) that the absolute value of the scattered magnetic field $|\varphi_s(\rho,\phi)|$ has the same profile as that of the $\varphi_s(\rho,\phi)$. 
\begin{figure}
\centering
\resizebox{0.4\textwidth}{!}
{\includegraphics{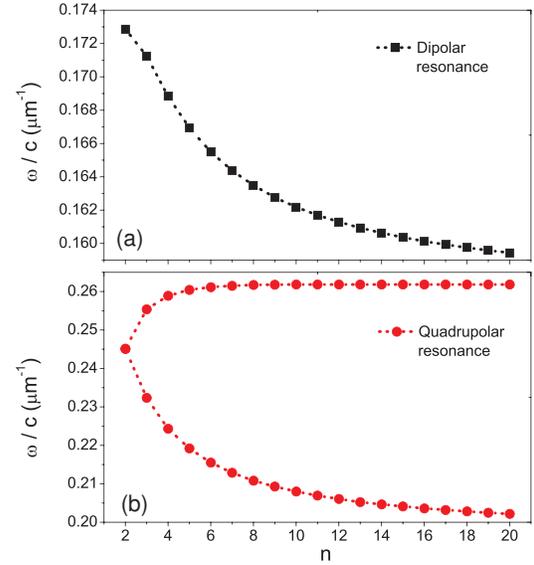}}
\caption{\label{fig:epsart} Resonant frequencies as a function of $n$ for quasi--square wire sections. All other parameters are the same as in Figure \ref{potencia_cuadrado_12}} \label{omega_dipolar} 
\end{figure}

In order to evaluate the dependence of the resonant frequencies on the $n$ parameter, in Figure \ref{omega_dipolar} we show the values of the frequency, indirectly estimated from the observation of positions of maxima of resonances in emission and scattering spectra, as a function of $n$, the other parameters as in Figure \ref{potencia_cuadrado_12}. 
We observe that when $n$ is increased, \textit{i.e.}, as the wire becomes more square, a significantly red shift of the dipolar resonant peak   (Figure \ref{omega_dipolar}a) and an increment in the splitting of the quadrupolar resonance (Figure \ref{omega_dipolar}b) occur. 

\begin{figure}
\centering
\resizebox{0.40\textwidth}{!}
{\includegraphics{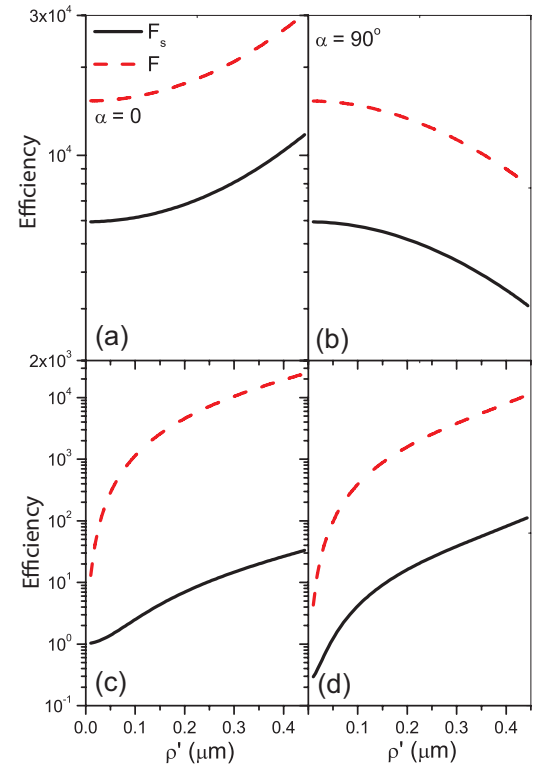}}
\caption{\label{fig:epsart} Emission  and radiation decay efficiencies (per unit length) as a function of the distance from the wire center for the dipole resonance frequency, $\alpha=0$ (a) and $\alpha=90^\circ$ (b), and for the quadrupole resonance frequency $\alpha=0$ (c) and $\alpha=90^\circ$ (d).  The emitter is located on the $\hat{x}$ axis and the resonance frequency is $\omega/c=0.1613\mu$m$^{-1}$ (a) and (b), $\omega/c=0.206\mu$m$^{-1}$ (c), $\omega/c=0.262\mu$m$^{-1}$ (d). All other parameters are the same as in Figure \ref{potencia_cuadrado_12}.
} \label{posicion_cuadrado}
\end{figure}
To evaluate the dependence of the emission and the radiation spontaneous decay rates with the location of the  source inside the quasi--square wire, in Figure \ref{posicion_cuadrado} we plot the efficiencies $F$ and $F_s$ as a  function of the distance $\rho'$ to the wire center. The dipole source is placed on the $\hat{x}$ axis.  In Figures  \ref{posicion_cuadrado}a and \ref{posicion_cuadrado}b we have chosen the emission frequency $\omega/c=0.1613\mu$m$^{-1}$, for which an enhancement of the emission and the radiation efficiencies occur due to the dipolar resonance excitation (first maximum observed in Figure \ref{potencia_cuadrado_12}). 
Unlike the circular case in which, for any location of the source, the values of the decay rates do not show any dependence on the orientation angle $\alpha$ \cite{cuevas2}, by comparing Figures  \ref{posicion_cuadrado}a and \ref{posicion_cuadrado}b  we see that, in the case of quasi--square wires, both $F$ and $F_s$ values are rather dependent on this angle.   
At $\rho'=0$ both the emission and the radiation decay rates are close to $10^4$ times larger than in the absence of the wire. These values are increasing as the source moves away from the wire center when the dipole axis is parallel to the $\hat{x}$ axis (Figure \ref{posicion_cuadrado}a). Conversely, when the dipole axis is parallel to the $\hat{y}$ axis, both $F$ and $F_s$ are  decreasing functions of $\rho'$  (Figure \ref{posicion_cuadrado}b). 
Figures \ref{posicion_cuadrado}c and \ref{posicion_cuadrado}d display the source location dependence  of the efficiency decay rates for the quadrupolar resonance frequencies, $\omega/c=0.206\mu$m$^{-1}$ (Fig. \ref{posicion_cuadrado}c) and $\omega/c=0.262\mu$m$^{-1}$ (Fig. \ref{posicion_cuadrado}d).  We can see that in both cases,  the efficiency values are similar and that these values are increasing functions of the distance to the wire. 

It is worth noting that the radiation  to  emission ratio, $F_s/F$ (quantum efficiency \cite{sanchezgil,sanchezgil2}), for the dipolar resonances (Figs. \ref{posicion_cuadrado}a and \ref{posicion_cuadrado}b) takes a value of approximately   $0.4$ regardless the position $\rho'$ of the dipole, a value that is comparable to those  obtained in other works where  simulations were made considering graphene based antennas at THz range \cite{filter,alu}. In addition, from these figures we see  that the 
 power radiated by the dipole inside the graphene wire is enhanced  $\approx 10^3-10^4$ times the  value  of the same dipole embedded in an unbounded medium.   
%
%

\begin{figure}
\centering
\resizebox{0.48\textwidth}{!}
{\includegraphics{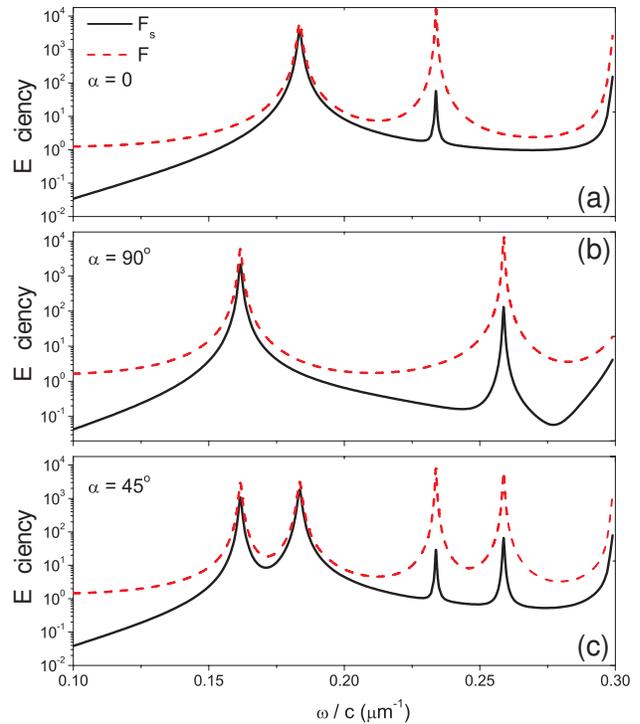}}
\caption{\label{fig:epsart} Efficiency (per unit length) curves,  calculated for a graphene coated quasi--rectangular wire with  $a=1.25\,b$ ($b=0.4037$) and $n=2$. The scaling factor $h$ is chosen so that the wire has the same perimeter as that of the circle in Figure \ref{comparacion}.  
The emitter is  localized at $\rho'=0.4\mu$m, $\phi'=180^\circ$ with its dipole moment oriented in the $+\hat{x}$ direction ($\alpha=0$) (a), in the $+\hat{y}$ direction ($\alpha=90^\circ$) (b), and $\alpha=45^\circ$ (c). All other parameters are the same as in Figure \ref{potencia_cuadrado_12}} \label{potencia_rectangulo}
\end{figure}

To further investigate the effect that the deviation from the circular geometry has on the emission spectrum, we consider  quasi--rectangles with $a=1.25\,b$.  In Figure \ref{potencia_rectangulo}, we plot the frequency dependence of the emission and radiation efficiencies for an emitter localized at $\rho'=0.4\mu$m and $\phi'=180^\circ$ (at $0.1\mu$m from the left of the wire). 
The break of the $90^\circ$ rotational symmetry of the wire section is  
manifested  in the dipolar resonance, which for the 
quasi--square case ($a=b$ and $n=12$) occurs near $0.16\mu$m$^{-1}$ and for the circular case ($a=b$  and $n=2$) occurs near $0.17\mu$m$^{-1}$, and that in the quasi--rectangular  case is split into two different resonant peaks, one near $0.1613\mu$m$^{-1}$ and the other near  $0.1835\mu$m$^{-1}$. The first peak corresponds to the dipole orientation $\alpha=90^\circ$ while the second peak corresponds to the dipole  orientation $\alpha=0$, as clearly indicated in Fig. \ref{potencia_rectangulo} by the fact that both resonances are decoupled for dipole directions parallel to either of the rectangle's axes and that the first (respectively second) peak is absent when the dipole direction is oriented along  (respectively perpendicular to) the major axis. 
\begin{figure}
\centering
\resizebox{0.50\textwidth}{!}
{\includegraphics{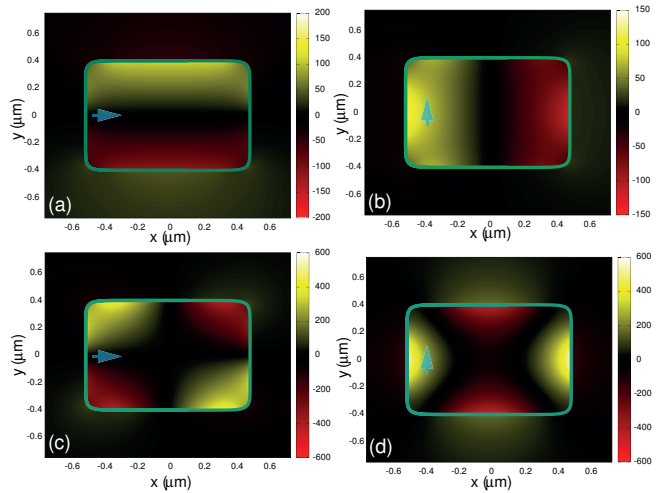}}
\caption{\label{fig:epsart} Map of the scattered magnetic field $\varphi_s(\rho,\phi)$ at a fixed time for plasmon resonances of the wire considered in Figure \ref{potencia_rectangulo}. Red, negative values, yellow, positive  values. The frequency emission is $\omega/c = 0.1835 \mu$m$^{-1}$ (a), $\omega/c = 0.1613 \mu$m$^{-1}$ (b), $\omega/c = 0.2339 \mu$m$^{-1}$ (c) and $\omega/c = 0.2588 \mu$m$^{-1}$ (d). The dipole emitter is indicated with an arrow.   
All parameters are the same as in Figure \ref{potencia_rectangulo}.} \label{campo_rectangulo}
\end{figure}

In Fig. \ref{campo_rectangulo} we plot the spatial distribution of the magnetic field for the wire considered in Fig. \ref{potencia_rectangulo}. In Fig. \ref{campo_rectangulo}a the frequency is $\omega/c=0.1613\mu$m$^{-1}$ and the dipole orientation is along the rectangle's minor axis  ($\alpha=90^\circ$) whereas in  Fig. \ref{campo_rectangulo}b the frequency is $\omega/c=0.1835\mu$m$^{-1}$ and the dipole orientation is along the rectangle's major axis ($\alpha=0$). 
We observe that in both cases the near field distributions follow the typical dipolar pattern, with two intensified field zones along the rectangle's major axis in Fig. \ref{campo_rectangulo}a or along the rectangle's minor axis in Fig. \ref{campo_rectangulo}b.  The field distributions in Figs. \ref{campo_rectangulo}c and \ref{campo_rectangulo}d,  calculated for $\omega/c=0.2339\mu$m$^{-1}$ and $\omega/c=0.2588\mu$m$^{-1}$, respectively, correspond to the quadrupole resonance frequencies splitting for which efficiency curves in Fig. \ref{potencia_rectangulo}c reach maximum values.

%
%

\section{Conclusions} \label{conclusiones}

In conclusion, we have examined the behavior of an optical emitter inside a graphene--coated subwavelength
wire of arbitrary cross section.  By using an electromagnetically rigorous integral  method, 
we have  investigated the modification of the emission and the radiation decay rates in the case of graphene  wires of quasi--rectangular cross section for varying the position and the dipole moment orientation of the emitter. To validate the method, we have compared the numerically computed efficiencies in the particular case of graphene--coated wires of circular section with the results obtained from an analytical theory.
In general, both the emission and the radiation efficiencies are strongly enhanced at frequencies where multipolar SP resonances are excited. The break of the rotational symmetry of the wire section leads to a frequency splitting of multipolar plasmonic resonances in the spectra for quasi--rectangular wires. Unlike the circular case, we found  a strong dependence of the efficiency spectra on the  dipole moment orientation.
We have shown that the multipolar order obtained  by the spectral position of the emission decay rate peak agrees well  with the multipolar order revealed by the topology of the near field.



\section*{Acknowledgment}
The author acknowledge the financial support of Consejo Nacional de Investigaciones Cient\'{\i}ficas y T\'ecnicas, (CONICET, PIP 451).

\end{document}